\documentclass[12pt]{article}
\usepackage{amsmath}
\usepackage{graphicx,psfrag,epsf}
\usepackage{enumerate}
\usepackage{natbib}
\usepackage{textcomp}
\usepackage[hyphens]{url} 
\usepackage{hyperref}
\providecommand{\tightlist}{%
  \setlength{\itemsep}{0pt}\setlength{\parskip}{0pt}}

\newcommand{\blind}{0}

\usepackage{color}
\usepackage{fancyvrb}

\DefineVerbatimEnvironment{Highlighting}{Verbatim}{commandchars=\\\{\}}
\usepackage{framed}
\definecolor{shadecolor}{RGB}{248,248,248}
\newenvironment{Shaded}{\begin{snugshade}}{\end{snugshade}}

\newcommand{\CommentTok}[1]{\textcolor[rgb]{0.56,0.35,0.01}{\textit{#1}}}

\newcommand{\DataTypeTok}[1]{\textcolor[rgb]{0.13,0.29,0.53}{#1}}
\newcommand{\DecValTok}[1]{\textcolor[rgb]{0.00,0.00,0.81}{#1}}

\newcommand{\KeywordTok}[1]{\textcolor[rgb]{0.13,0.29,0.53}{\textbf{#1}}}
\newcommand{\NormalTok}[1]{#1}
\newcommand{\OperatorTok}[1]{\textcolor[rgb]{0.81,0.36,0.00}{\textbf{#1}}}

\newcommand{\StringTok}[1]{\textcolor[rgb]{0.31,0.60,0.02}{#1}}

\usepackage{booktabs}
\usepackage{longtable}
\usepackage{array}
\usepackage{multirow}
\usepackage{wrapfig}
\usepackage{float}
\usepackage{colortbl}
\usepackage{pdflscape}
\usepackage{tabu}
\usepackage{threeparttable}
\usepackage{threeparttablex}
\usepackage[normalem]{ulem}
\usepackage{makecell}
\usepackage{xcolor}

\usepackage{xcolor, soul, xspace, float, subfig}

\begin{document}

\def\spacingset#1{\renewcommand{\baselinestretch}%
{#1}\small\normalsize} \spacingset{1}


\if0\blind
{
  \title{\bf ``Playing the whole game'': A data collection and analysis exercise with
Google Calendar}

  \author{
        Albert Y. Kim \thanks{Albert Y. Kim is Assistant Professor, Statistical \& Data Sciences,
Smith College, Northampton, MA 01063 (e-mail:
\href{mailto:akim04@smith.edu}{\nolinkurl{akim04@smith.edu}}). This work
was not supported by any grant. The authors thank numerous colleagues
and students for their support.} \\
    Program in Statistical \& Data Sciences, Smith College\\
     and \\     Johanna Hardin \\
    Department of Mathematics, Pomona College\\
      }
  \maketitle
} \fi

\if1\blind
{
  \bigskip
  \bigskip
  \bigskip
  \begin{center}
    {\LARGE\bf ``Playing the whole game'': A data collection and analysis exercise with
Google Calendar}
  \end{center}
  \medskip
} \fi

\bigskip
\begin{abstract}
We provide a computational exercise suitable for early introduction in
an undergraduate statistics or data science course that allows students
to `play the whole game' of data science: performing both data
collection and data analysis. While many teaching resources exist for
data analysis, such resources are not as abundant for data collection
given the inherent difficulty of the task. Our proposed exercise centers
around student use of Google Calendar to collect data with the goal of
answering the question `How do I spend my time?' On the one hand, the
exercise involves answering a question with near universal appeal, but
on the other hand, the data collection mechanism is not beyond the reach
of a typical undergraduate student. A further benefit of the exercise is
that it provides an opportunity for discussions on ethical questions and
considerations that data providers and data analysts face in today's age
of large-scale internet-based data collection.
\end{abstract}

\noindent%
\vfill

\newpage
\spacingset{1.45} 

\hypertarget{introduction}{%
\section{Introduction}\label{introduction}}

The title of our paper refers to the reality that in order to master a
subject, one needs to do more than practice the individual, elemental,
and necessary parts. For example, no matter how good you are at running,
dribbling the ball, or shooting the ball into the upper corner of the
net, you cannot excel at soccer unless you practice \textbf{the whole
game} \citep{perkins2010making}. While \citet{wickham2019pckg} use the
phrase to describe creating an entire R Package from beginning all the
way to distribution on GitHub, we use the term to describe the entire
process by which data science is performed.

\begin{figure}[H]

{\centering \includegraphics[width=0.8\linewidth]{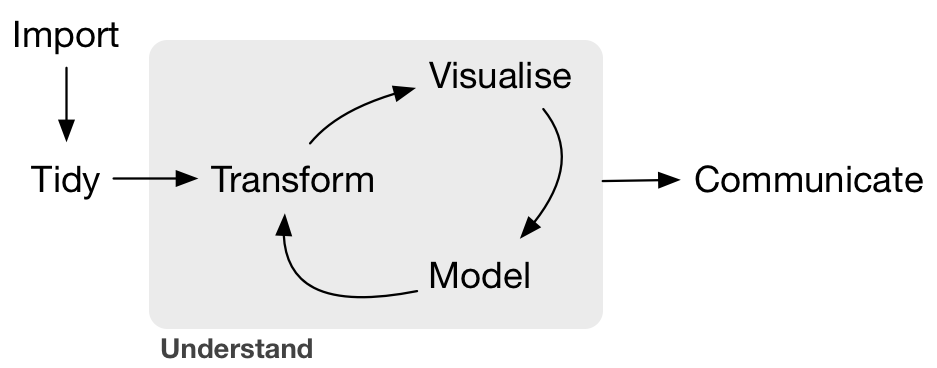} 

}

\caption{Data science workflow diagram \citep{wickham2017r}}\label{fig:data-science-pipeline}
\end{figure}

Many statistics and data science educators are likely quite familiar
with \citet{wickham2017r}'s model of the tools needed in a typical data
science project seen in Figure \ref{fig:data-science-pipeline}. While
one should not interpret any simplifying diagram in a overly literal
fashion, we appreciate two aspects of this diagram. First, the cyclical
nature of the ``Understand'' portion emphasizes that in many substantive
data analysis projects, original models and visualizations need
updating, necessitating many iterations through this cycle until a
desired outcome can be communicated. Second, it encourages a holistic
view of the elements of a typical data science project.

Previously many undergraduate statistics courses have not taught such a
holistic approach, instead focusing only on individual components at the
expense of others, thereby leaving gaps in the process. In this paper,
we view these gaps in terms of what \textbf{Not So Standard Deviations}
podcast hosts Roger Peng and Hilary Parker call differences between
``what data analysis is'' and ``what data analysts do''
\citep{NSSD-smoothie}. In other words, there are many differences
between the idealized view of the data analysis process and what is
actually done in practice.

We argue that is important to adjust curricula to ``bridge the gaps''
between ``what data analysis is'' and ``what data analysts do''
\citep{McNamara2015}. For example, a substantial bridge of the gap has
come through great pedagogical strides to expose students in statistics
and data science courses to the entirety of the process in Figure
\ref{fig:data-science-pipeline} \citep[\citet{hardin2015ds},
\citet{loy2019}, \citet{yan2019}]{baumer2015ads}.

\begin{figure}[H]

{\centering \subfloat[Image credit: \citep{ruiz2014infoworld}\label{fig:80wrang1}]{\includegraphics[width=0.49\linewidth]{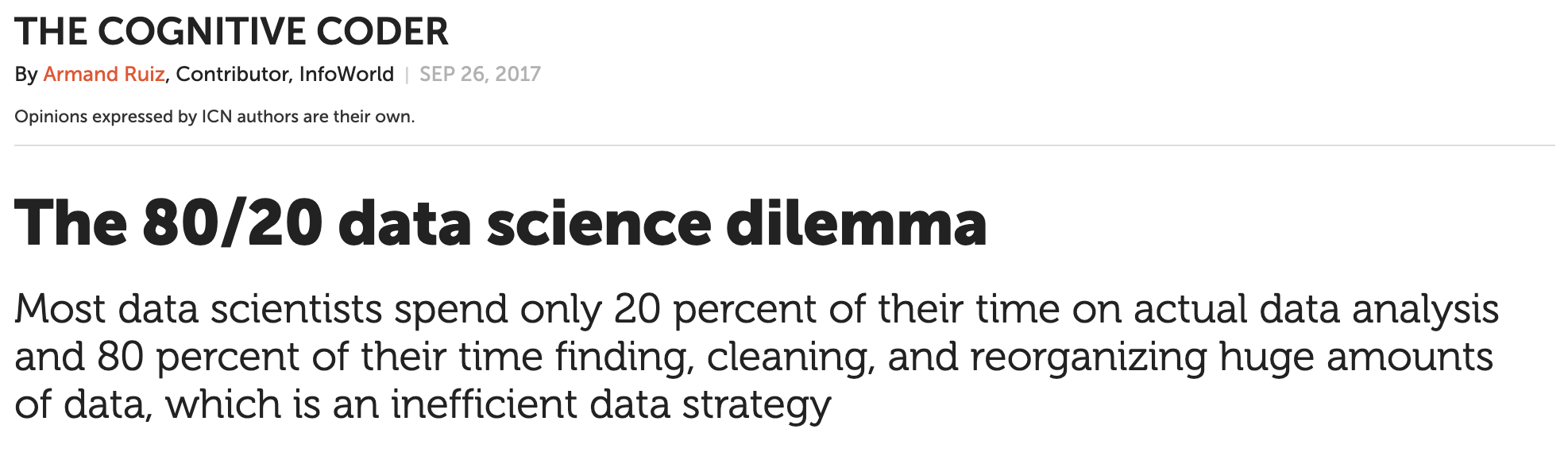} }\subfloat[Image credit: \citep{drob2018twitter}\label{fig:80wrang2}]{\includegraphics[width=0.49\linewidth]{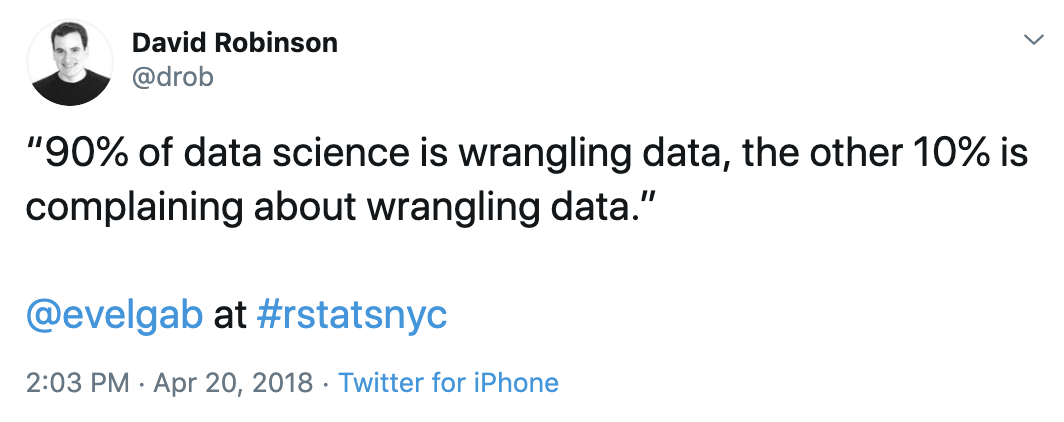} }

}

\caption{The 80-20 rule of data wrangling}\label{fig:80wrang}
\end{figure}

Another example of an existing gap in some statistics and data science
courses is sparse treatment of data wrangling. A familiar refrain from
working data scientists is that 80\% of their time is spent wrangling
data leaving only 20\% for actual analysis (see Figure
\ref{fig:80wrang}). Given the prospect of the 80-20 rule,
\citet{kim2018tise} argue that to completely shield students in
statistics courses from performing meaningful data wrangling is to do
them a disservice. \citet{horton2015chance} propose five key data
wrangling elements that deserve greater emphasis in the undergraduate
curriculum: creative \& constructive thinking, facility with different
types of data, statistical computing skills, experience wrangling, and
an ethos of responsibility. Two of the most seminal contributions to
modernizing the statistics and data science curriculum, \citet{NASEM}
and \citet{Nolan2010}, both emphasize the importance of teaching and
practicing data wrangling.

In this manuscript, we argue that another such gap in curricula is
inadequate treatment of the data collection component of the cycle. Many
working data scientists do not spend a large proportion of their time
working on one static data set as suggested by the single ``Import''
step in Figure \ref{fig:data-science-pipeline}
\citep[\citet{Lohr2014}]{ruiz2014infoworld}. Instead, the data used to
address questions are often dynamic; the data change based on
differences in time, sampling strategies, questions asked, variables
collected, etc. Furthermore, only after a first iteration of data has
been collected and analyzed can research questions be updated and fine
tuned, often necessitating another round (or more) of data collection.

\begin{figure}[H]

{\centering \includegraphics[width=1\linewidth]{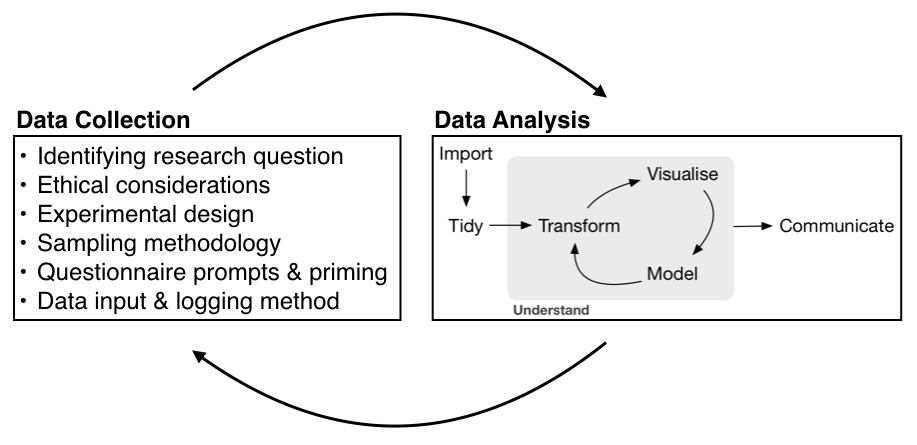} 

}

\caption{``Playing the whole game": Data collection as well as data analysis in data science.}\label{fig:whole-game}
\end{figure}

While we are not the first to argue that data collection and acquisition
is an important topic for the statistics classroom \citep{Zhu2013} or
the computer science classroom
\citep[\citet{Protopapas2020}]{Blitzstein2013}, we point out that
\citet{wickham2017r}'s model of data analysis from Figure
\ref{fig:data-science-pipeline} is not a complete representation of a
typical ``data science project'' as it neglects a critical phase:
(repeated) data collection. We present what we term ``playing the whole
game'' in Figure \ref{fig:whole-game}, which augments the earlier ``data
analysis'' diagram in Figure \ref{fig:data-science-pipeline} with an
additional ``data collection'' block. The new block comprises of key
elements to consider when collecting data, including ethical
considerations, the experimental design (if any), the sampling
methodology, questionnaire design in the case of surveys, the data input
and logging methods used, and most critically, identifying the research
question. While the new block is by no means exhaustive and thus should
not be interpreted in an overly literal fashion either, we again
emphasize the cyclical and iterative nature of data collection and data
analysis.

Note that some of the tasks in the ``Data Collection'' box may not be
seen immediately as collecting data. However, we believe that they are
important considerations at the data collection step. For example as it
relates to ``Ethical Considerations'', questions like ``Should you
collect someone's location data?'' or ``Should you require explicit
permission to collect a particular variable?'' would be considered. As
it relates to ``Question Prompts'', studies on the psychological effect
of ``priming'' have shown that you can steer survey responses to a
particular direction depending on how and when questions are asked
\citep{Hjortskov2017}.

In the context of undergraduate statistics and data science classrooms,
a large amount of data analysis conducted by students uses data that
they themselves or the class as a whole had no part in collecting. For
example, many students seek data for assignments and projects from
repositories like Kaggle.com or data.gov. Thus in many cases, students
get no exposure to the ``data collection'' component of Figure
\ref{fig:whole-game} and take the data for granted. Lack of exposure to
data collection has potential pitfalls including potentially erroneous
conclusions based on mistaken assumptions of the data collection method,
a lack of a sense of ownership of the work, and most importantly, no
experience going through the iterative process of ``playing the whole
game.''

Previous literature has focused on slightly different ways of
incorporating data collection into the classroom: research projects
(e.g., \citet{Halvorsen2000}, \citet{Halvorsen2010}, \citet{Sole2017});
simulating realistic data through games (e.g., \citet{Kuiper2015}); and
collecting classroom activity data (e.g., \citet{Scheaffer2004}).

\begin{figure}[H]

{\centering \includegraphics[width=0.7\linewidth]{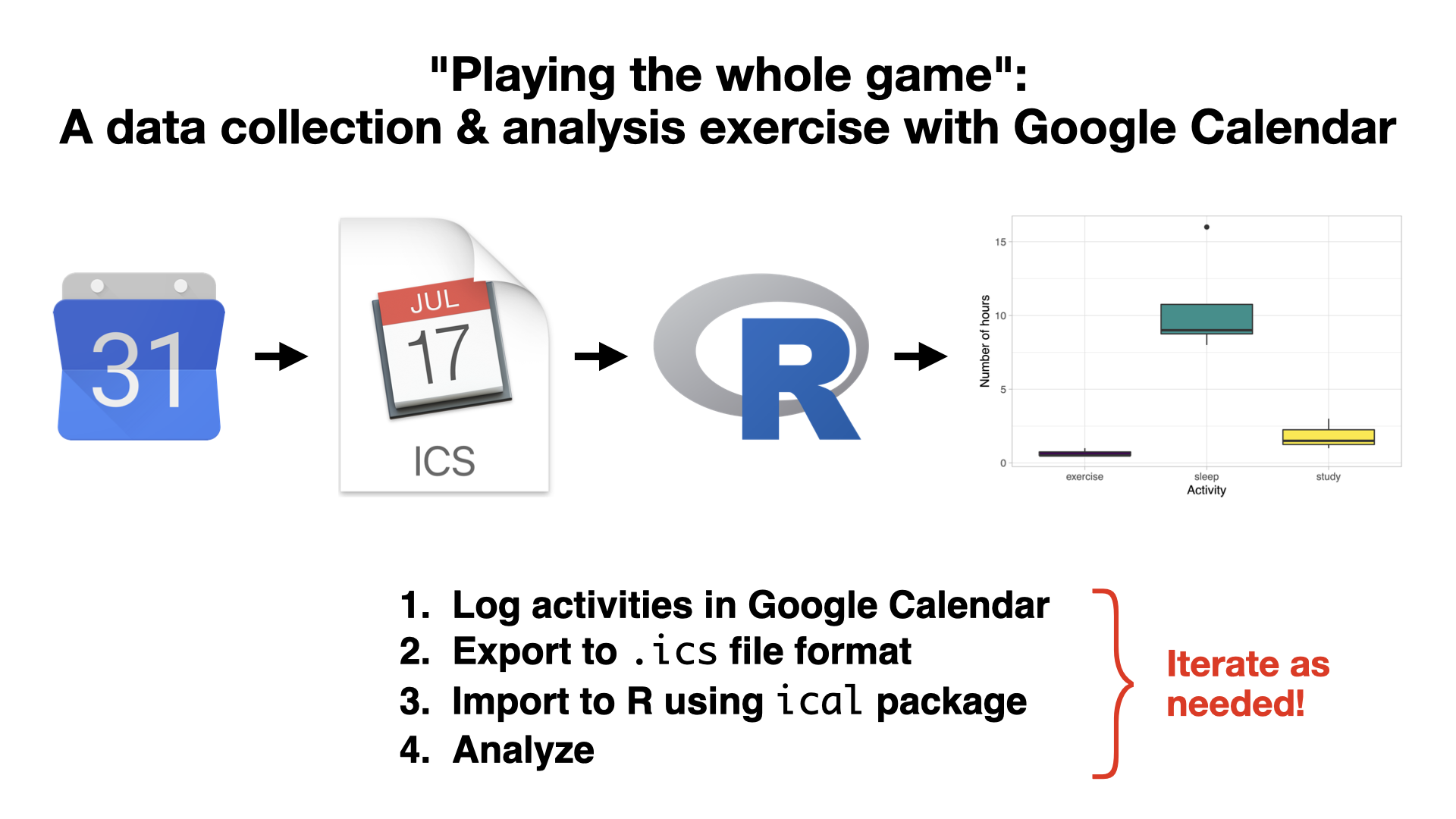} 

}

\caption{Graphical representation of playing the ``Playing the Whole Game" with Google Calendar.}\label{fig:whole-game-calendar}
\end{figure}

Building toward the goal of having students experience the entire data
analysis process, we propose a data collection activity that allows
students to mimic the real-life data collection conducted by numerous
internet-age organizations in industry, media, government, and academia.
At the same time the technological background and sophistication
necessary for the activity is kept at a level suitable for undergraduate
classroom settings. The classroom activity touches many components of
playing the whole game, demonstrates to the students the difficulty of
data science, and provides an interesting research question with which
the students can engage. A visual summary of the activity is presented
in Figure \ref{fig:whole-game-calendar}.

\hypertarget{muse}{%
\subsection*{Muse}\label{muse}}
\addcontentsline{toc}{subsection}{Muse}

The idea for our ``data collection'' exercise for students came out of
an episode of the earlier mentioned \textbf{Not So Standard Deviations}
podcast titled ``Compromised Shoe Situation'' \citep{NSSD-shoe} (as well
as described in a corresponding blog post by Roger Peng titled ``How
Data Scientists Think - A Mini Case Study'' \citep{peng2019blog}).

Hosts Hilary Parker and Roger Peng gave each other a data science
challenge whereby they had to solve a problem using data science. They
contrasted it to other common data science challenges (such as the
American Statistical Association's DataFest \citep{datafest} or
prediction and classification competitions available from Kaggle.com)
where the data have already been collected and cleaned.

The challenge that Parker and Peng proposed centers on identifying what
factors influence the time it takes each of them to get to work. In one
of our favorite discussions of their podcast, they break down the
iterative process of gathering data, learning what information it
provides, and gathering more data. They also discuss the difficulty of
gathering precise information and the balance of ``low-touch'' and
``high-touch'' data collection (collection methods that require
\emph{few} active user actions versus \emph{many}, respectively). In
Parker's data collection, she implemented a low-touch system of
recording when her commute started by automatically recording when her
phone disconnects from her home WiFi; a higher-touch part of her
analysis was when she manually logged the route she took to work.

In the rest of the manuscript we describe how we have taken the ideas
from Parker and Peng's podcast and infused them into a class assignment
previewed in Figure \ref{fig:whole-game-calendar} which allows students
to practice ``playing the whole game.'' In Section \ref{sec:MM} we
provide the details of the assignment and the considerations that went
into why we made some of the choices we did. Section \ref{sec:results}
provides a reflection on what we learned and how the assignment
succeeded in accomplishing the goals for ``playing the whole game.'' In
Subsection \ref{quotes}, we include quotes from reflection pieces
written by students summarizing their experiences. As a vital aspect of
our work, in Section \ref{sec:discussion} we discuss the ethical
considerations of the assignment and how they can be generalized to the
larger scale data collection done by internet-age organizations in
industry, media, government, and academia. As part of our ethics
discussion, we emphasize the importance of bringing up data ethics all
along the semester with every data analysis, and not just in a data
ethics course.

\hypertarget{sec:MM}{%
\section{Materials \& Methods}\label{sec:MM}}

\hypertarget{context}{%
\subsection{Context}\label{context}}

In Fall 2019 both authors were each teaching classes in a context where
the whole game would be important to the learning outcomes of the
course. Albert's class was ``Introduction to Data Science'' at Smith
College, a no-prerequisite course designed to appeal to a broad audience
and act as an entry-point to the Statistical and Data Sciences major. He
gave the assignment in the fifth week of the semester as the first
``mini-project.'' Jo's class was ``Computational Statistics'' at Pomona
College with primarily junior and senior math/statistics majors; she
used the assignment as the first homework of the
semester.\footnote{The work has been verified by the Smith College Institutional Review Board as "Exempt" according to 45CFR46.101(b)(1): (1) Educational Practices.}

In what follows, we provide the details of the assignment. To engage the
students maximally, we tried to find a question with universal appeal
and landed on ``How do I spend my time?'' In the assignment, the student
is required to collect, wrangle, visualize, and analyze data based on
entries they make in their own electronic calendar/planner application,
such as Google Calendar, macOS, or Microsoft Outlook.

\begin{figure}[H]

{\centering \includegraphics[width=0.6\linewidth]{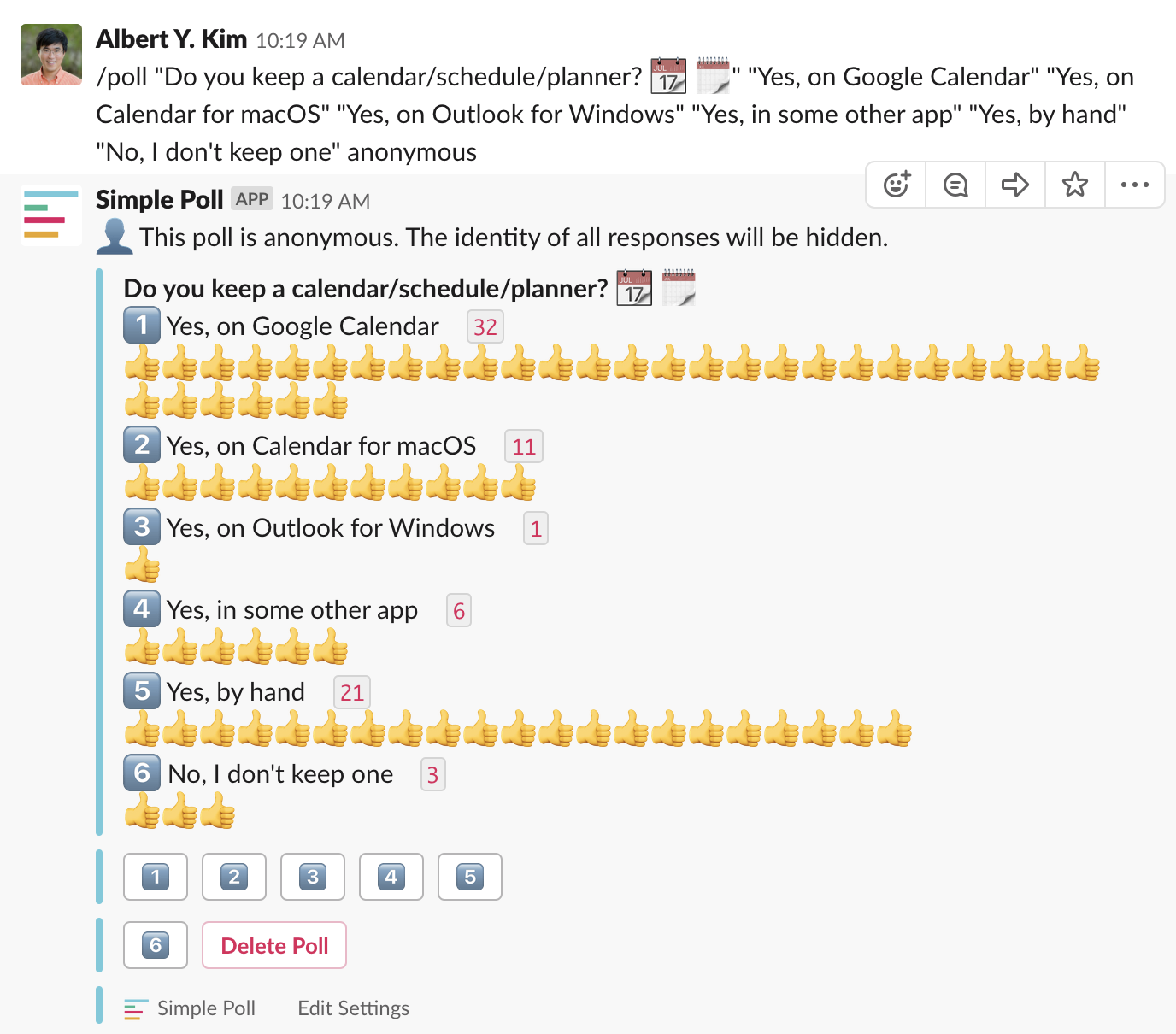} 

}

\caption{Poll: Do you keep a calendar/schedule/planner?}\label{fig:class-poll}
\end{figure}

To assess prior student use of a calendar/planner, Albert polled his
class at Smith College on their preferred method of keeping track of
their schedule (See Figure \ref{fig:class-poll}). In Albert's case, most
students were already keeping an electronic calendar (44 of 74), while
some were keeping a hand-written calendar (21 of
74).\footnote{Both authors were surprised to see that any college student would be able to make it through the semester without any type of calendar at all (3 of 74)!}

\hypertarget{sec:LO}{%
\subsection{Learning Outcomes}\label{sec:LO}}

The learning outcomes for the exercise can be broken down into three
categories. The first category is ``Data collection'' which focuses on
the technical aspects of the activity. The second category is ``Data
ethics'' which focuses on the larger context and conclusions resulting
from the exercise. The last category encompasses the learning goals
related to ``Playing the whole game.''

\begin{itemize}
\tightlist
\item
  Data collection

  \begin{enumerate}
  \def\labelenumi{\arabic{enumi}.}
  \tightlist
  \item
    Experience creating measurable data observations (e.g., how is ``one
    day'' measured, or what defines ``studying'').
  \item
    Address data collection constraints due to limits in technological
    capacity and human behavior.
  \end{enumerate}
\item
  Data ethics

  \begin{enumerate}
  \def\labelenumi{\arabic{enumi}.}
  \setcounter{enumi}{2}
  \tightlist
  \item
    Practice the ethical and legal responsibilities of those collecting,
    storing, and analyzing data.
  \item
    Decide limits for personal privacy.
  \item
    Deliberate on the trade-offs between research results and privacy.
  \end{enumerate}
\item
  ``Playing the whole game''

  \begin{enumerate}
  \def\labelenumi{\arabic{enumi}.}
  \setcounter{enumi}{5}
  \tightlist
  \item
    Tie together data collection, analysis, ethics, and communication
    components.
  \item
    Iterate between and within the components of the ``whole game.''
  \end{enumerate}
\end{itemize}

Throughout the manuscript, we touch on all of the learning outcomes.
Section \ref{sec:benact} describes many of the benefits and learning
outcomes related to ``Data collection''; Section \ref{sec:bengame}
describes many of the benefits and learning outcomes related to
``Playing the whole game''.

As a specific example of learning outcome 5, consider ethics training
for animal studies where there is typically a discussion on the minimum
number of samples needed for ``ethical research.'' The idea is that an
under-powered study would indicate that the animals had been needlessly
sacrificed with no possibility of moving scientific research forward. As
with minimum number of samples, we ask the students what types of
research connect with which types of privacy violations, and where is
the correct balance for pushing forward knowledge. A more complete
examination of data ethics in the classroom is given in Section
\ref{sec:discussion}.

\hypertarget{the-assignment}{%
\subsection{The Assignment}\label{the-assignment}}

We have provided the complete assignments used by both authors at the
following website
\url{https://smithcollege-sds.github.io/sds-www/JSE_calendar.html/}.
While the two versions of the assignment vary slightly in format and
reflection, both still require the students to go through the entire
process of collecting, wrangling, visualizing, analyzing, and
communicating about the data, with an important additional step of
reflecting on the information gathered. The assignment was scaffolded so
that a student with minimal computational and coding experience could
still directly work with calendar data.

Before we go into the details of working with calendar data, we first
point out some differences in approach between the two assignments. One
unique aspect of Albert's class is that he had the students work in
pairs. A particular student would make their calendar entries and then
export the data. However, instead of analyzing their own calendar, they
would send their data to their partner who then wrangled, visualized,
and analyzed the data (as well as vice versa). The motivation was to
encourage students to think about what data they could or should share
with their partner, what data they couldn't nor shouldn't, and any
particular responsibilities that the individual analyzing the data had.

In Jo's class, she spent half of one class period discussing the podcast
in great detail. Because it was early on in the semester, there were
lots of unknowns about what is data science and what are the possible
ways one can use data to make decisions. Additionally, the assignment
and podcast continued to come up in class sessions throughout the
semester as a way to ground conversations with respect to data
collection \& analysis: what we can do, what we can't do, what we should
do, and what we shouldn't do.

\hypertarget{details-of-working-with-calendar-data}{%
\subsection{Details of Working With Calendar
Data}\label{details-of-working-with-calendar-data}}

Students were instructed to track how they spent their time on their
calendar application for approximately 10-14 days. More specifically,
students were instructed to fill in blocks of time and mark the entry
with the activity they were performing: sleeping, studying, eating,
exercising, socializing, etc. The students chose their own categories to
fit their own schedules. Students were able to have overlapping blocks
of activities in situations where two activities were done
simultaneously. Students were also informed that they should feel
comfortable leaving out any details they did not want to share; indeed,
if they wanted to, they were free to make up all the information in the
calendar.

Note that our example centers around the use of Google Calendar. To
ensure as much consistency as possible we encouraged students who did
not have an electronic calendar to use Google Calendar to record their
activities. However, the assignment can equally be done using macOS
calendar or Microsoft Outlook.

As an example, we filled a sample Google Calendar with entries between
September 2nd and 7th, 2019, which you can view using the Google
Calendar interface at \url{http://bit.ly/dummy_calendar}. We suggest
that after scrolling to the week of September 1st, 2019, you click the
``Week'' tab on the top right for a week-based overview of the calendar
entries.

\begin{figure}[H]

{\centering \includegraphics[width=0.7\linewidth]{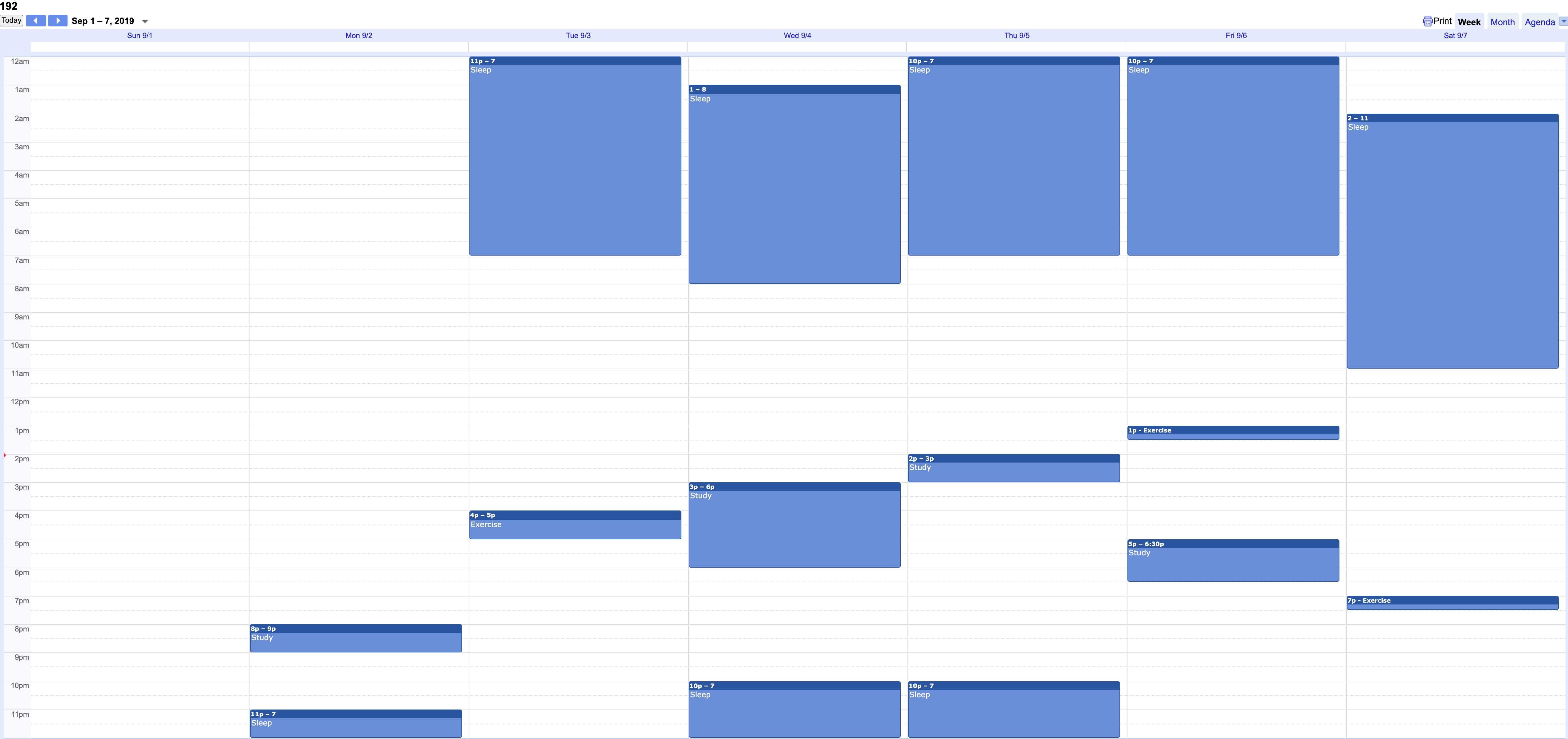} 

}

\caption{Sample Google Calendar presented to students.}\label{fig:calendar}
\end{figure}

After looking at the sample calendar, students exported their own
calendar data to \texttt{.ics} file format, a universal calendar format
used by several email and calendar programs, including Google Calendar,
macOS Calendar, and Microsoft Outlook. Their file was then imported into
R as a data frame using the \texttt{ical\_parse\_df()} function from the
\texttt{ical} package \citep{R-ical}.

In order to help the students focus on the larger data science paradigm,
the instructors scaffolded the assignment in template R Markdown files
\citep{R-rmarkdown} which served as the foundation of the students'
submissions; we provide these scaffolded assignments at
\url{https://smithcollege-sds.github.io/sds-www/JSE_calendar.html/}.
Continuing our earlier example involving a sample Google Calendar, we
exported its contents to a \texttt{192.ics} file. Subsequently, we
imported the \texttt{.ics} file into R as a \texttt{tibble} data frame
\citep{R-tibble}, and then performed some data wrangling using
\texttt{dplyr} \citep{R-dplyr} and the \texttt{lubridate} package for
parsing and wrangling dates and times \citep{R-lubridate}.

Below, we give an extract of the code representing the crux of the
process previously described. (Note that to change the
\texttt{"America/New\_York"} timezone setting above, run the
\texttt{OlsonNames()} function in R to output a list of all time zones.)

\begin{Shaded}
\begin{Highlighting}[]
\KeywordTok{library}\NormalTok{(ggplot2)}
\KeywordTok{library}\NormalTok{(dplyr)}
\KeywordTok{library}\NormalTok{(lubridate)}
\KeywordTok{library}\NormalTok{(ical)}

\NormalTok{calendar_data <-}\StringTok{ "192.ics"} \OperatorTok{%>%}\StringTok{ }
\StringTok{  }\CommentTok{# Use ical package to import into R:}
\StringTok{  }\KeywordTok{ical_parse_df}\NormalTok{() }\OperatorTok{%>%}\StringTok{ }
\StringTok{  }\CommentTok{# Convert to "tibble" data frame format:}
\StringTok{  }\KeywordTok{as_tibble}\NormalTok{() }\OperatorTok{%>%}\StringTok{ }
\StringTok{  }\CommentTok{# Use lubridate packge to wrangle dates and times:}
\StringTok{  }\KeywordTok{mutate}\NormalTok{(}
    \DataTypeTok{start_datetime =} \KeywordTok{with_tz}\NormalTok{(start, }\DataTypeTok{tzone =} \StringTok{"America/New_York"}\NormalTok{),}
    \DataTypeTok{end_datetime =} \KeywordTok{with_tz}\NormalTok{(end, }\DataTypeTok{tzone =} \StringTok{"America/New_York"}\NormalTok{),}
    \DataTypeTok{length =}\NormalTok{ end_datetime }\OperatorTok{-}\StringTok{ }\NormalTok{start_datetime,}
    \DataTypeTok{date =} \KeywordTok{floor_date}\NormalTok{(start_datetime, }\DataTypeTok{unit =} \StringTok{"day"}\NormalTok{)}
\NormalTok{    ) }\OperatorTok{%>%}
\StringTok{  }\CommentTok{# Make calendar entry summary all lowercase:}
\StringTok{  }\KeywordTok{mutate}\NormalTok{(}\DataTypeTok{summary =} \KeywordTok{tolower}\NormalTok{(summary)) }\OperatorTok{%>%}\StringTok{ }
\StringTok{  }\CommentTok{# Compute total daily time spent on each activity:}
\StringTok{  }\KeywordTok{group_by}\NormalTok{(date, summary) }\OperatorTok{%>%}
\StringTok{  }\KeywordTok{summarize}\NormalTok{(}
    \DataTypeTok{length =} \KeywordTok{sum}\NormalTok{(length),}
    \DataTypeTok{length =} \KeywordTok{as.numeric}\NormalTok{(length)}
\NormalTok{    ) }\OperatorTok{%>%}
\StringTok{  }\CommentTok{# Set correct time interval length units. Here, hours:}
\StringTok{  }\KeywordTok{mutate}\NormalTok{(}\DataTypeTok{hours =}\NormalTok{ length}\OperatorTok{/}\DecValTok{60}\NormalTok{) }\OperatorTok{%>%}\StringTok{ }
\StringTok{  }\CommentTok{# Filter out only rows where date is later than 2019-09-01:}
\StringTok{  }\KeywordTok{filter}\NormalTok{(date }\OperatorTok{>}\StringTok{ "2019-09-01"}\NormalTok{)}
\end{Highlighting}
\end{Shaded}

\begin{figure}[H]

{\centering \includegraphics[width=0.8\linewidth]{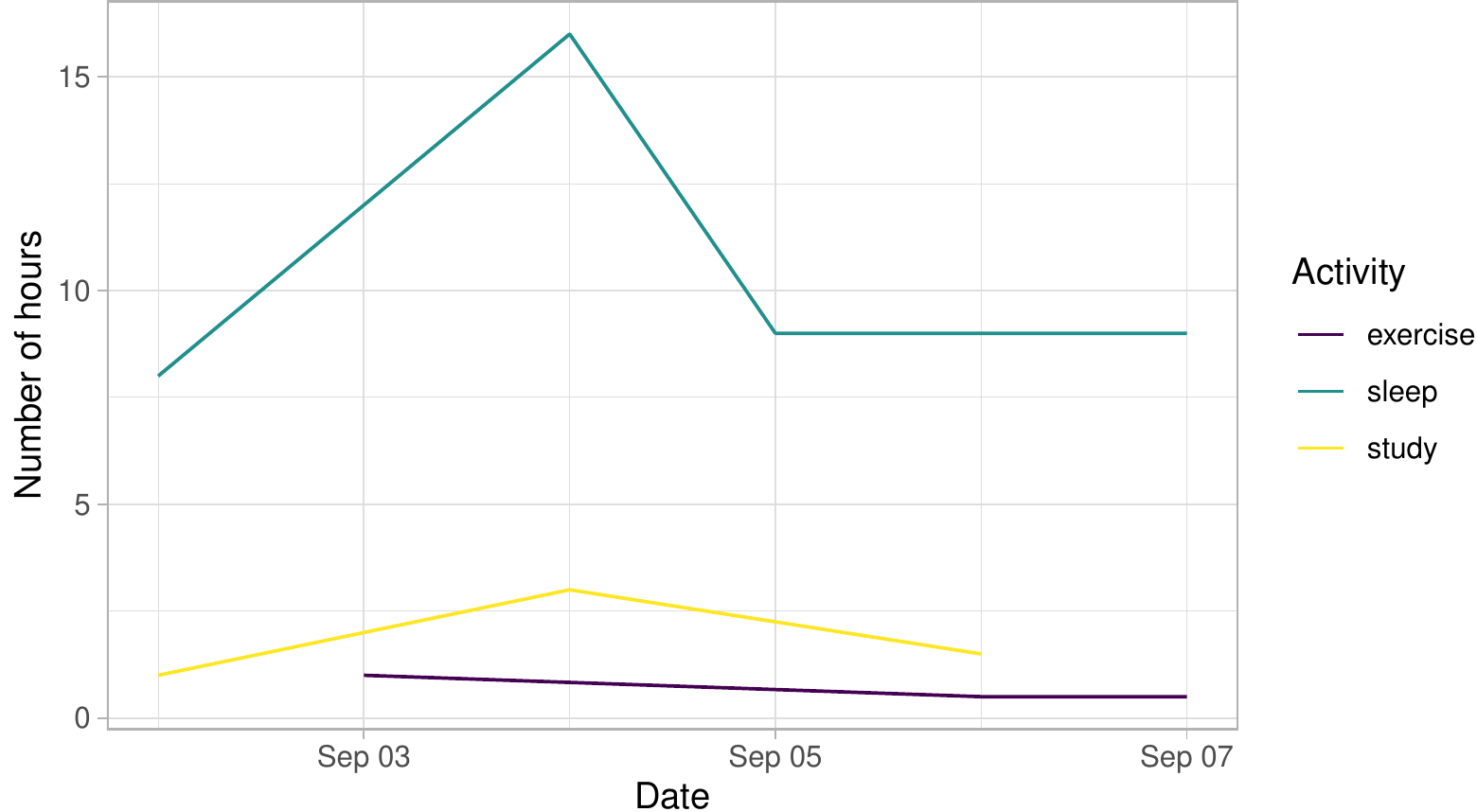} 

}

\caption{Time series plot of calendar data.}\label{fig:example-calendar-plot-1}
\end{figure}

\begin{figure}[H]

{\centering \includegraphics[width=0.8\linewidth]{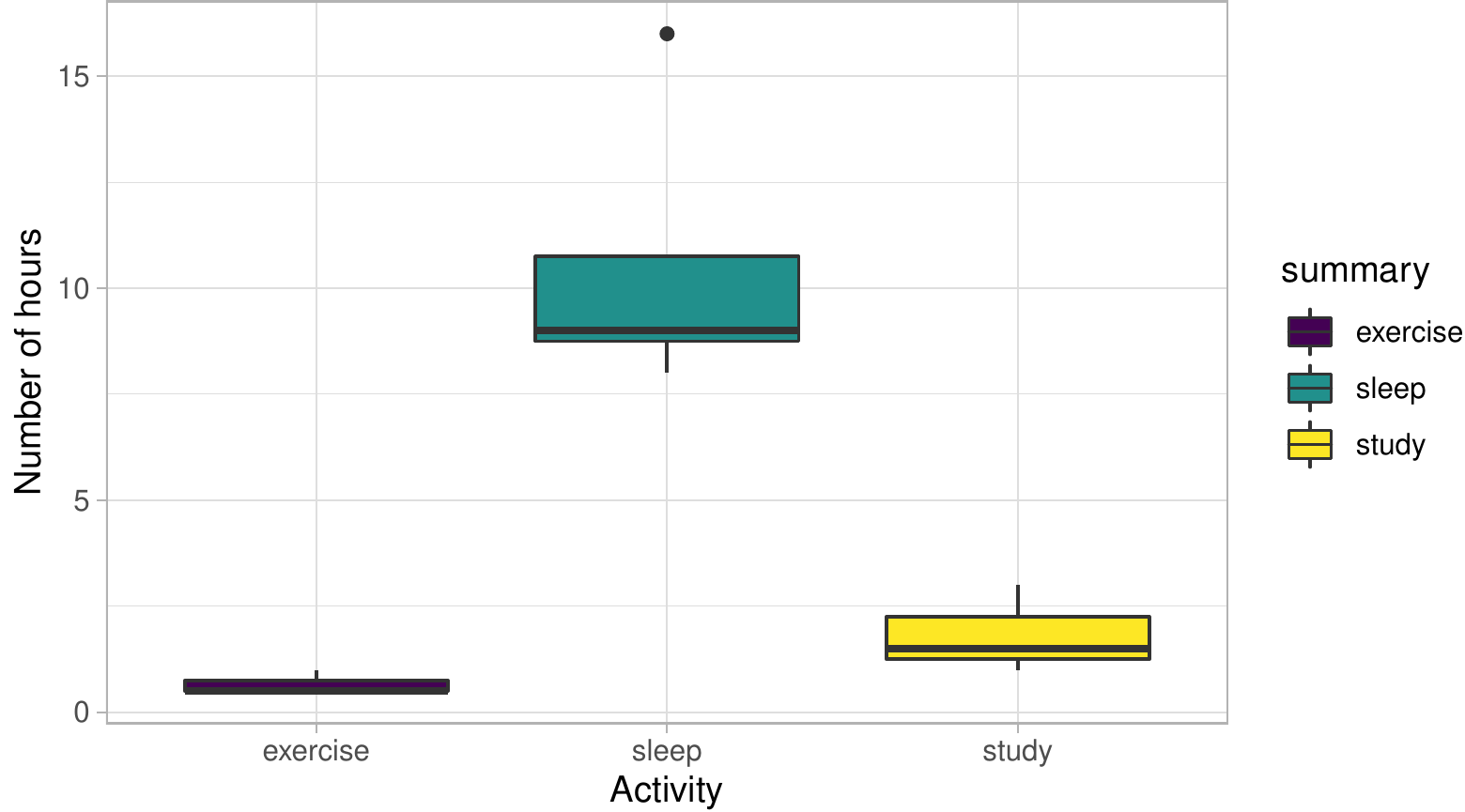} 

}

\caption{Distribution of number of hours spent split by activity type.}\label{fig:example-calendar-plot-2}
\end{figure}

The resulting \texttt{calendar\_data} data frame output is presented in
Table \ref{tab:example-calendar-data}. The \texttt{calendar\_data} data
frame can then be used to make plots like the time-series linegraph in
Figure \ref{fig:example-calendar-plot-1} and the side-by-side boxplot in
Figure \ref{fig:example-calendar-plot-2}.

\begin{table}[!h]

\caption{\label{tab:example-calendar-data}Example calendar data frame.}
\centering
\begin{tabular}[t]{llrr}
\toprule
date & summary & length & hours\\
\midrule
2019-09-02 & sleep & 480 & 8.0\\
2019-09-02 & study & 60 & 1.0\\
2019-09-03 & exercise & 60 & 1.0\\
2019-09-04 & sleep & 960 & 16.0\\
2019-09-04 & study & 180 & 3.0\\
2019-09-05 & sleep & 540 & 9.0\\
2019-09-06 & exercise & 30 & 0.5\\
2019-09-06 & study & 90 & 1.5\\
2019-09-07 & exercise & 30 & 0.5\\
2019-09-07 & sleep & 540 & 9.0\\
\bottomrule
\end{tabular}
\end{table}

\hypertarget{sec:results}{%
\section{Results}\label{sec:results}}

The assignments we used in our classes did not take a substantial amount
of time to create, execute, or discuss. Albert engaged with the
assignment for about half a class period at three different times: (1)
to motivate and introduce the assignment including an end-to-end
demonstration; (2) a check-in to make sure everyone was on track, and
(3) an in-class discussion after the assignment had been submitted. Jo
spent only a few minutes discussing the technical aspects of the
assignment and 30 min during one class period discussing
\citet{NSSD-shoe}. We believe that including the ``play the whole game''
activity in the course allowed for a much deeper understanding and
reflection on the entire data science pipeline, without needing to
substitute any of our previous course learning goals. That is to say,
the assignment we have shared is in keeping with the current structure
of many statistics and data science courses. In fact, while our
manuscript was under review, Prof.~Katharine F. Correia at Amherst
College extended our original ideas and modified them so that her
students could create ``data diaries'' during the social isolation
period of Spring 2020. Some examples of her students' remarkable work
can be found here:
\url{https://stat231-01-s20.github.io/data-science-diaries/}.

We now discuss what we view as the successes and lessons of the activity
(for student perspectives on these successes and lessons see Section
\ref{quotes}).

\hypertarget{sec:benact}{%
\subsection{Benefits Inherent to the Activity}\label{sec:benact}}

First, the focus on ``self'' and an engaging question (``How do I spend
my time?'') easily captured our students' interest. Universally, they
found the assignment compelling, with one student remarking informally
afterwards that they planned on pursuing further analyses on their own
private Google Calendar. Another student came up with a clever idea of
documenting their intended time studying versus their actual time
studying. Some students seemed concerned that their question on how they
used their time wasn't interesting enough, but we emphasized that the
project should be more about process than outcome. Additionally, we
prompted students with follow-up queries such as: ``Do the amounts of
time spent on certain activities have an inverse relationship with each
other?'' or ``Do you spend as much time on a particular activity as you
expect to?'' (where the latter question would necessitate recording both
expected and actual time for each event).

Second, having the data collection involve time intervals puts a nice
cap/standardization on the scope on the type of data collected. As can
be seen in Table \ref{tab:example-calendar-data} of the example
\texttt{calendar\_data} data frame, all resulting data frames consist of
three variables: date (numerical), length of time spent (numerical), and
activity type (categorical). Having such a combination of variables is
an excellent starting point for data science courses. Furthermore, many
students encountered very interesting computational and data wrangling
questions quite organically, some along the lines of:

\begin{itemize}
\tightlist
\item
  How do I combine two types of activities in R?
\item
  How do I only select events from days past 2019-09-02?
\item
  My `hours spent' variable is actually in minutes rather than hours.
  How do I convert it to hours? (We noticed operating system level
  variation in the default units of the recorded \texttt{length}
  variable in the \texttt{calendar\_data} dataframe.)
\item
  How would I make a scatterplot of daily time spent studying versus
  time spent sleeping?
\end{itemize}

The latter question is particular interesting as it necessitates
converting data frames from tall/long format (like that of the
\texttt{calendar\_data} example in \ref{tab:example-calendar-data}) to
wide format using the \texttt{pivot\_wider()} function from the
\texttt{tidyr} package \citep{R-tidy}. In our opinions, this question
serves as an ideal method to teach the subtle concept of ``tidy'' data
\citep{wickham2014jss}. Rather than starting with the somewhat technical
definition of ``tidy'' data, there is a benefit to having students
gently ``hit the wall'' whereby they cannot create a certain
visualization or wrangle data in a certain way unless the data is in the
correct format.

Lastly, both instructors found that the exercise served as an excellent
opportunity to teach students about file paths and including the dataset
(in this case the .ics file) in an appropriate directory. Students were
required to compile .Rmd files which connected to a unique .ics file,
and a second individual (either a partner or a student grader) also
needed to be able to compile the .Rmd file. Due to the typical structure
of a homework assignment (e.g., professor providing a link to a
dataset), our experience is that working with file paths is both
difficult for novice data science students as well as difficult to
provide examples for practicing.

\hypertarget{sec:bengame}{%
\subsection{Benefits from ``Playing Whole Game''}\label{sec:bengame}}

Importantly, there were benefits engendered by having students ``play
the whole game,'' both (1) of a technical, coding, and computational
nature and (2) of a scientific and research nature. Many lessons were
only learned by students after having gone through an entire iteration
of ``data collection'' and ``data analysis'' cycle as seen in Figure
\ref{fig:whole-game}.

For example, one student made calendar entries recording the start times
of events, whereas the assignment assumed time intervals were being
recorded. For example, ``I went to sleep at time X.'' versus ``I slept
between times X and Y, and thus slept for Z hours.'' The student thus
needed to revise their data collection method.

Other students found out that unless their calendar entry titles matched
exactly, they are recorded as two different activity types. Consider the
following extremely nuanced difference due to trailing spaces:
\texttt{"Sleep"} versus \texttt{"Sleep\ "}, such data collection methods
were also revised in order to ensure entries were standardized. These
students thus learned about the differences in how humans and computers
process text data.

Another issue that surprised the instructors was that calendar entries
based on a ``repeat'' schedule only appeared once in their resulting
data in R: the entry for the first day. The instructors thus had to
consider whether to update the assignment instructions for students in
future classes, or to let them encounter this issue on their own.

On top of more technical, coding, and computational issues, sometimes
students realized mid-way that their research question itself needed
revising and thus had to retroactively fill in their calendars using
their best guesses. These students thus realized the importance of
iterating through the entire process sooner rather than later in order
to prevent errors from accumulating.

An unforeseen issue that combined both technical and scientific queries
was how to allocate sleep hours. For example, if a student went to sleep
at 10pm and woke up at 6am, then 2 hours of sleep would be logged for
one day and 6 hours for the next, whereas most people would call it a
single night of sleep.

\hypertarget{quotes}{%
\subsection{Student quotes}\label{quotes}}

On top of the earlier mentioned student ``data diaries'' from
Prof.~Katharine F. Correia's course available at
\url{https://stat231-01-s20.github.io/data-science-diaries/}, we now
provide excerpts from the reflection pieces students wrote in Albert's
course at Smith College. Students were asked to write a joint reflection
piece on their experiences, keeping the podcast in mind
\citep{NSSD-shoe}.\footnote{Per Smith College Institutional Review Board guidelines, explicit consent was obtained directly from all students who are confidentially quoted in the manuscript.}
In particular:

\begin{itemize}
\tightlist
\item
  As someone who provides data: What expectations do you have when you
  give your data?
\item
  As someone who analyzes other's data: What legal and ethical
  responsibilities do you have?
\end{itemize}

The student quotes have been grouped into categories which relate back
(roughly) to the previously described learning outcomes (LO) in
Subsection \ref{sec:LO}.

\textbf{Relating to iterating between ``data collection'' and ``data
analysis'' (LO1)}:

\begin{itemize}
\tightlist
\item
  ``I fell ill and was unable to log in more data points for her to
  visualize. Student X was still able to visualize my activity,
  incorporating my sick days as another variable''
\item
  ``One of the technical issues we ran into, and probably the defining
  experience of this project, was the difficulty in creating consistent
  error free information to export to our partner. It's kind of
  ridiculous how small manual entry errors, like whether or not I used
  spaces (in the calendar entries), made such a difference at the end of
  the line when mystery variables started showing up''
\item
  ``But that was not the biggest issue we encountered, our issue was the
  fact that we inputted repeated events on our calendar which affected
  how many observations came up onto the data set (since only the first
  of the multiple events would show up)''
\item
  ``I, for example, learned that data is fickle and needs consideration
  of how variables interact with each other within a data visualization
  before unnecessary data collection. Her original question was
  inadequate as one of her variables wasn't a function of time; thus,
  she had to retroactively enter the data for her new question using her
  best estimate, meaning her data should be recognized as imprecise and
  that it could affect the outcome of any analysis.''
\end{itemize}

\textbf{Relating to ``low'' and ``high touch'' data collection methods
(LO2)}:

\begin{itemize}
\tightlist
\item
  ``Specifically for Student X's data, it was interesting to work with
  real-time data recorded by her phone. Because it was automatically
  collected by her mobile device, there was less room for human errors
  when calculating the time spent on her phone screen.''
\item
  ``The time is hard to pin down because sometimes there will be
  interruptions or potentially interleaving studying to do other things.
  We think that the study time could be off between an hour and a half
  an hour for regular weekdays and maybe two during weekends because
  sometimes it is harder to take down the exact start and stop time on
  the spot. Comparatively, recording sleeping time is much easier
  because we could always check the time when the alarm goes off and
  take a screenshot when setting the alarm for the other day.''
\end{itemize}

\textbf{Relating to analyst ethical responsibilities (LO3)}:

\begin{itemize}
\tightlist
\item
  ``For example, in our group, one member accidentally shared their
  entire calendar data with the other member when only a small portion
  of this data was needed for analysis. The other member realized the
  first member's mistake, deleted the data, and showed the first member
  how to separate the events she wanted to be analyzed from the rest of
  her calendar.''
\end{itemize}

\textbf{Relating to prior analyst biases (LO3)}:

\begin{itemize}
\tightlist
\item
  ``It might be worthwhile as the analyzer to be completely transparent
  and address any previous biases that may affect the reliability of a
  given data analysis.''
\item
  ``Keep to the original data, don't change the data to get a favorable
  result''
\end{itemize}

\textbf{Relating to data provider expectations (LO4)}:

\begin{itemize}
\tightlist
\item
  ``Additionally, the individual should have the right to ask about the
  research project that the data is going to and how their data may be
  used.''
\end{itemize}

\textbf{Relating to analyst empathy towards data providers (LO5)}:

\begin{itemize}
\tightlist
\item
  ``The question was straightforward but it also made us uneasy as we
  had to consider both what we were comfortable sharing and how to
  handle our partner's data in an ethical manner''
\item
  ``On the other hand, because of my discomfort, I was also able to
  handle Student X's data with the same caution that I would expect her
  to do with my dataset''
\item
  ``This in turn motivated me to handle her data with care and
  sensitivity. For this reason, I appreciated this project as it forced
  us to simultaneously consider our roles both as people who share our
  data in numerous scenarios throughout our daily lives and as data
  science students with a new responsibility to handle another person's
  data ethically.''
\item
  ``If we oppose Facebook tracking our personal information and
  profiting off it without our consent, we owe it to our
  customers/providers of data to not do the same.''
\item
  ``Personally I did question how comfortable I was sharing information
  about XXX, but decided to push my boundaries, because though it made
  me nervous to share information about XXX I was also very curious
  about how it affected YYY, and the actual risk of sharing this
  information was little to none. I wanted to gain some insight from
  this project that would help me lead a healthier, more stable life.''
\end{itemize}

\textbf{Relating to the value of data (LO5)}:

\begin{itemize}
\tightlist
\item
  ``There would be no incentives for us to share our data if we are not
  getting anything back from it.''
\item
  ``A grey area of data analysis would be targeted ads. On the one hand,
  one could argue that getting targeted ads bring convenience to only
  getting advertised products we would enjoy. On the other hand, one
  could also argue it pushes capitalism on us and thus only
  disadvantages the public by heavily influencing us to buy products we
  might not need.''
\end{itemize}

\textbf{Relating to the balancing act many analysts face (LO5)}:

\begin{itemize}
\tightlist
\item
  ``Information that puts our security or identity at risk should be
  left out. However, if too little information is shared, it may affect
  the ability to accurately represent any phenomena occurring within the
  data.''
\item
  ``In the podcast we listened to, Hilary Parker specifically mentioned
  how in collecting employees' commute time, she wanted to protect
  privacy by withholding the exact minute they left work. An alternative
  was releasing the general length of time employees spend commuting to
  work would, but this would remove the impact of confounding variables
  and externatlities. The exact times an employee commuted were
  important because part of the data collection included factoring in
  traffic patterns. Because of this consideration, Parker decided it
  would be more effective to round the times an employee arrived of left
  work to the nearest hour to maximize privacy and accuracy of the
  data.''
\end{itemize}

\textbf{Relating to actionable insight (LO6)}:

\begin{itemize}
\tightlist
\item
  ``Since then I have been more cautious about granting apps access to
  my information.''
\item
  ``Visualization 1 made me realize that I should also spend more time
  during the weekdays sleeping and hanging out with friends to take care
  of myself.''
\end{itemize}

\textbf{Relating to revised research questions (LO7)}:

\begin{itemize}
\tightlist
\item
  ``For example, while analyzing Student X's data, we had to decide
  whether or not Friday would be considered a weekday or a weekend.''
\item
  ``As for improvements in our next projects, for consistency and
  accuracy, it might be important to keep the data collection dates
  fixed so that there are the same number of the week days considered in
  the analysis.''
\end{itemize}

\hypertarget{sec:discussion}{%
\section{Discussion}\label{sec:discussion}}

As noted by \citet{baumer2020}, many major professional societies,
including the American Statistical Association (ASA) \citep{asa-ethics},
the Association for Computing Machinery (ACM) \citep{acm2018ethics}, and
the National Academy of Sciences (NAS) \citep{national2009being}, have
long published guidelines for conducting ethical research. However, only
more recently has there been growing literature on the weaving of data
ethics topics early and often all throughout undergraduate curricula in
data science, statistics, and computer science \citep[\citet{ottSDSS},
\citet{burton2018teach}, \citet{elliott2018teaching},
\citet{heggeseth-sdss}]{baumer2020}.

The class activity we present provides a myriad of ways to start a
classroom conversation about data ethics. Albert motivated the idea of
the project by discussing the pros and cons of how he logs his daily
calorie consumption using the MyFitnessPal app. On the one hand, the
students could appreciate the convenience of logging calories using the
barcode scanner, on the other hand, they could also understand that
Albert is essentially telling the Under Armour corporation everything he
eats and when. In the age of Fitbit, 23andMe, and innumerable different
health and fitness apps, the example presents the students with an
opportunity to think about what information they are sharing with large
corporations.

Instructors may choose a different type of data collection, or let each
student decide on their own collection method. Some possible extensions
include the above mentioned fitness apps (outputting all entries in a
.csv file); every interaction on Instagram (outputting a .json file);
Google trends or Google search terms (output as a function of time); or
personal Facebook information (as .html or .json). Each data type will
present a unique setting in which to discuss privacy and data ethics
considerations.

In Jo's class conversation about the podcast, one student appreciated
that Hilary and Roger discussed whether they \emph{should} do the data
collection. The student expressed that maybe Hilary should just leave
earlier for work so as to not make anyone wait. That would be the more
kind thing to do. Indeed, as Hilary herself says ``\ldots{} it's
probably better to just give yourself a cushion so you don't have to
worry.'' The student comment led to a larger conversation about what
types of questions we should answer with data science and what types of
questions should not be data driven.

In the last few minutes of the podcast, Parker says \citep{NSSD-shoe}:

\begin{quote}
\ldots{} setting up, thinking through what could you collect, what makes
sense to collect? What are the assumptions in the data that you're
collecting? Like going through that, I would encourage people to go
through that process, even if you end up like not doing it long-term or
something, but it's still, I think that's the skill set that's important
to develop if you want to work in data science.
\end{quote}

Why does Hilary Parker implore people to ``play the whole game?'' We
believe that there is a disconnect in the narrative surrounding ``what
is data analysis'' when instead it should be more closely connected to
``what do data analysts do.'' The only way to inculcate lessons
pertaining to ``what data analysts do'' is by having students ``play the
whole game.'' Furthermore, just as in real-life, it is critical that
students go through the cycle several times. In other words, ``iterate
early and iterate often.''

\hypertarget{acknowledgements}{%
\section{Acknowledgements}\label{acknowledgements}}

Most importantly, we'd like to acknowledge the Smith College and Pomona
College students with whom we work. It is from their ideas and energy
that we discovered the fun and beauty of working in statistics and data
science. Hilary Parker \& Roger Peng provided not only the inspiration
for the activity but also helpful feedback on the pre-print. We
appreciate the thoughtful suggestions of three anonymous referees and
the associate editor. On top of the packages cited in the manuscript,
the authors also used the \texttt{ggplot2} \citep{R-ggplot2},
\texttt{kableExtra} \citep{R-kableExtra}, \texttt{knitr}
\citep{R-knitr}, and \texttt{viridis} packages \citep{R-viridis}. We
also thank Hadley Wickham for providing the direct insipiration for the
``playing the whole game'' wording in our paper.

\hypertarget{declaration-of-interest-statement}{%
\section{Declaration of Interest
Statement}\label{declaration-of-interest-statement}}

Nothing to declare.

\bibliographystyle{agsm}
\bibliography{bibliography.bib}

\end{document}